# CAstelet in Virtual reality for shadOw AVatars (CAVOAV)

Georges Gagneré[1], Anastasiia Ternova[2]

[1] Scènes du monde, création, savoirs critiques (EA 1573), Paris 8 University, France
[2] INREV AIAC (EA 4010), Paris 8 University, France

**Corresponding author:** Georges Gagneré, georges.gagnere@univ-paris8.fr

**Abstract**
After an overview of the use of digital shadows in computing science research projects with cultural and social impacts and a focus on recent researches and insights on virtual theaters, this paper introduces a research mixing the manipulation of shadow avatars and the building of a virtual theater setup inspired by traditional shadow theater (or "castelet" in french) in a mixed reality environment. It describes the virtual 3D setup, the nature of the shadow avatars and the issues of directing believable interactions between virtual avatars and physical performers on stage. Two modalities of shadow avatars direction are exposed. Some results of the research are illustrated in two use cases: the development of theatrical creativity in mixed reality through pedagogical workshops; and an artistic achievement in *The Shadow* performance, after H. C. Andersen.

## 1. Introduction

The essence of performing art is an stimulating issue when exploring the potentialities of virtual reality on a theatrical stage : virtual actors compete with living performers. Working with shadows is very helpful because they payed an outstanding tribute to digital art and they used to guide humans in supernatural worlds.

This paper details an applied variation of AvatarStaging theatrical framework (Gagneré & Plessiet, 2018) in the wake of traditional shadow theater and dreams on virtual theater. Bringing together the immemorial shadow power to make us live incredible stories and the emergent fascinating expressive potentialities of virtual reality, we propose a mixed reality theatrical setup aimed at fostering the dialog between physical and digital entities. We call it CAstelet in Virtual reality for shadOw AVatar (CAVOAV) as an English translation for the French acronym "CAstelet Virtuel d'Ombre AVatar".

Section 2 is a short survey on digital shadows in computing science research projects with cultural and social impacts and gives examples of virtual theater research projects. Section 3 describes the specificities of CAVOAV derived from the AvatarStaging framework concerning the set design and the nature of the shadOw AVatars ("Ombre AVatar" in French, OAV) populating it. It explains how the virtual OAVs are lively animated. Section 4 offers an illustration with two applied use cases: the first is realized in a pedagogical context, and the second is a professional performance that toured in France and abroad. Section 5 concludes and gives some perspectives.

## 2. Related works

Shadows, that materialize the presence of an object or a body in a stream of light by the absence of light impact on the surrounding space, have always fascinated human beings. In the Allegory of the Cave, in his work *Republic*, Plato used them to illustrate how imperfect is our perception of the real world. In his *Natural History*, Pliny explained how the clay modeler Butades discovered the art of painting by seeing his daughter drawing the outlines of her lover's head shadow cast by a lantern. More widely, shadows inspired a great variety of myths and tales amid all the peoples and gave birth to numerous pieces of art and literature.

In the child cognitive development, understanding shadows marks from 5-6 years old the appropriation of distant action concept. In the history of sciences itself, shadows have played a key role in many discoveries, including that of the peripheral place of the Earth in the solar system, which overturned worldviews in the Renaissance (Casati & Cavanagh, 2019).

It is therefore not surprising that shadows are to be found at the core of the first interactive art experiences in a mixed environment linking a physical participant and a computer generated virtual reality, which were led by (Krueger & al., 1985) within the VIDEOPLACE system in the mid 1970's. With this artistic installation, Krueger built the first virtual theater with digital silhouette shadows allowing the onlooker to interact with an intelligent artificial environment. After these first experiences, capturing the silhouette shadow with video cameras in visible or infrared light remained a widespread way of integrating a participant in an interactive digital installation (Jacquemin, Gagneré & Lahoz, 2011) (Obushi & Koshino, 2018).

(Pasquier, Han, Kim, & Jung, 2008) proposed the concept of Shadow Agent to make human-machine interactions more natural, especially in a dialogue with artificial agents. (Jacquemin, Ajaj & Planes, 2011) listed the various instantiations of digital shadows for a virtual dancer circulating between real and virtual spaces. (Batras, Jégo, Guez & Tramus, 2016) also used shadows to let a participant play with a virtual actor capable of improvising. Thus, using shadows seems to be a promising way for allowing a physical performer to act in a virtual environment.

Besides, the progress of low-cost motion capture devices from the 2010s, such as the Microsoft Kinect camera or inertial motion capture suits (for example Noitom's Perception Neuron since 2015), have fostered the exploration of virtual reality as a possible space for a virtual theater. (Wu, Boulanger, Kazakevich & Taylor, 2010) introduced a real-time performance system for virtual theater. (Jdid, Richir & Lioret, 2013) detailed the characteristics of virtual reality which would offer the spectator a renewed immersive experience. (Pietroszek, Eckhardt & Tahai, 2018) demonstrated an actor management tool in a computational live theater. CAVOAV system aims to bring together these two tracks of research by proposing a new instantiation of digital shadows within a virtual theater in a mixed reality theatrical environment.

## 3. CAVOAV Setup Architecture

The setup is derived from the AvatarStaging mixed reality framework (Gagneré & Plessiet, 2018).

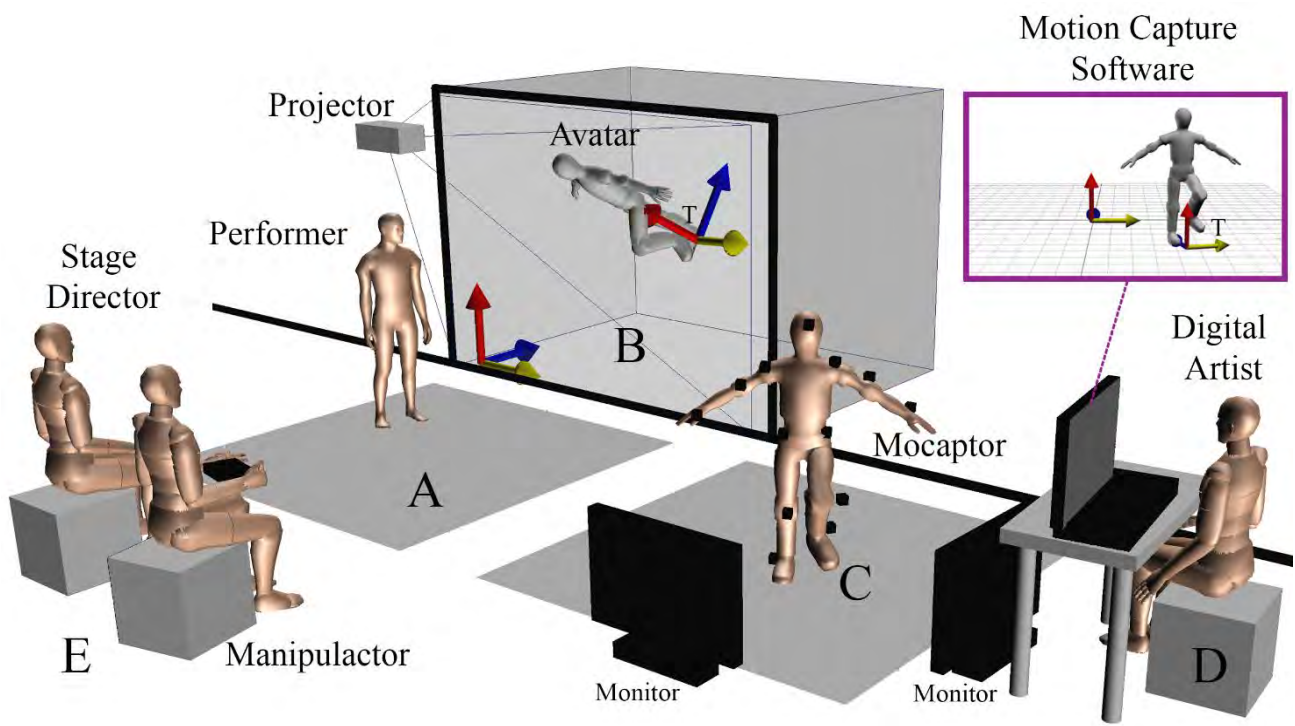

**Figure 1: AvatarStaging mixed reality framework.**

To control the avatar, the mocaptor wears a motion capture suit and the manipulactor uses a gamepad.

In AvatarStaging global framework, a performer wearing a motion capture suit (that we called mocaptor) acts in space C and controls an avatar in virtual space B interacting with a physical performer in space A, in front of B. A and B form the mixed reality stage in front of the audience E. His partner using hand devices as gamepad or midi controller (that we called manipulactor) helps the mocaptor control his avatar in the mixed reality stage.

CAVOAV setup specifies the nature of both the set design and the avatars in this way:
1) Virtual space B represents a traditional theatrical stage behind a frame,
2) Avatars are flat human silhouette shadows with 2 types of behavior.
It uses two different approaches for directing the avatars in the virtual set.

### 3.1. Set Design Specificity

In the expressions "shadow theater" or "puppet theater", the word theater has two meanings. The first is the style of living art using shadow or puppet which is different in nature from the performing art with physical actors. It requires either immaterial silhouettes or manipulated objects. The second meaning is the specific architecture necessary to present shadows or puppets to the audience. The traditional structure is often small and raised, that allows puppeteers to easily move behind the frame and manipulate several puppets. Moreover, the structure is equipped with light and set design to produce shadow effects or set changes. In French, this small theatrical structure, that could remind of a large model, is named "castelet".

The idea of the virtual "castelet" is to simulate both the living art style of playing with shadows and puppets, and the physical structure with its light and set requirements. CAVOAV is the French acronym for the "CAstelet Virtuel d'OmbrAVatar" concept, that we adapt in English as CAstelet in Virtual reality for shadOw AVatar.

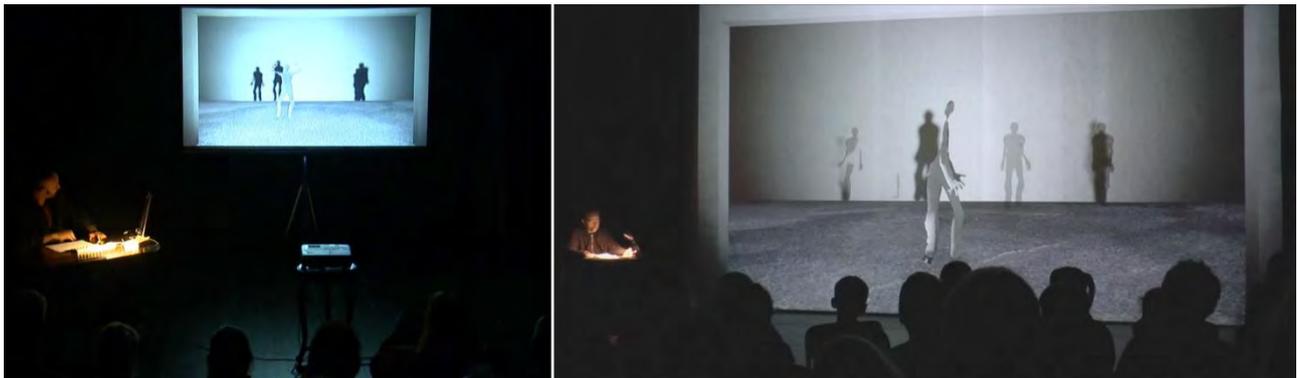

**Figure 2 : Shadow Theater and Shadow Avatars**

Left: the video projected virtual space, with the size of a traditional puppet theater, is beside the performer; right: the virtual space is slightly bigger that human size and in continuity with the physical stage.

Fig. 2 shows two versions of the setup. Left image shows the performer sit left side behind a small table with a desk lamp on. The virtual stage is projected on a kind of elevated slide screen, approximatively the same size as a traditional shadow theater. The performer reads a text to the audience and interacts with a shadow avatar looking at him from the screen. Right image shows the same situation differently arranged: the screen is much bigger and seems to give access to a real stage with strange actors. The set inside the virtual stage is composed of a backstage translucent screen letting appear "traditional" shadows that seem to come from characters on the other side and lit from behind. It exemplifies one of the numerous combinations allowed by the setup virtual nature to play both with shadows and/or puppets characters.

### 3.2. Nature and shadowing behavior of the shadOw AVatar (OAV)

In fig.2 right, an OAV stands front stage looking back at the translucent screen. It is a flat human silhouette moving in 3D, inspired from the *Peter Schlemihl's Miraculous Story* by Chamisso (1814) in which the Devil peels from the floor Peter Schelmihl's shadow cast by the sunlight, and keeps it in exchange of a magic bottomless gold sack that makes fortune to his owner. An OAV is therefore a sort of flat shadow able to move in 3D. In the physical world, it could be figured out by a piece of paper cut with a human shape, but it surely won't be easily manipulated. In a video game engine, it is easily built by rigging any human flat shape and controlling it through a motion capture device.

Of course, an OAV projects its own shadow (see fig. 4). However, this « natural » shadow of the frontstage OAV is invisible because its light rendering feature is set to "non casting shadow", which is a common property in 3D software. Anyway, an example of this shadow is visible on the backstage translucent screen. It seems that OAVs move backstage behind the translucent screen and are lit with a rear projector, that produces the traditional result that one sees in shadow theater. For rendering reasons, instead of being produced by a rear light, the shadows are made up with an "invisible" OAV in front of it, using the 3D graphic object feature not to be rendered, but keeping on its feature "casting a shadow". They are therefore invisible but cast a shadow. Numerous other graphic combinations can be explored in the setup with this OAV avatar. One of the main motivations for CAVOAV is to play with the idea of circulation between different states of "reality". From the state of a shadow on the wall, the OAV accesses a 3D existence as if one peels its shadow. This transformation is visible on the two OAVs at each extremity of the translucent screen in fig. 2b. By giving the possibility to precisely simulate a realistic shadow effect meanwhile developing it in a magical direction by common 3D visual effects, the virtual theater offers a powerful tool to foster the suspension of disbelief necessary to bring out dramatic situations.

### 3.3. OAV direction

For building a virtual theater, after the set and the characters, the stage direction is one of the key issues. Following Plessiet classification (Plessiet, Gagneré & Sohier, 2019), it would be great to cast virtual actors that can play autonomously on stage, or at least virtual golems that could follow by themselves the director instructions. These actors are however out of our scope for the moment, and we propose to direct OAVs in the 3D space as virtual puppets according to two approaches.

3.3.1. Living and animated virtual puppets

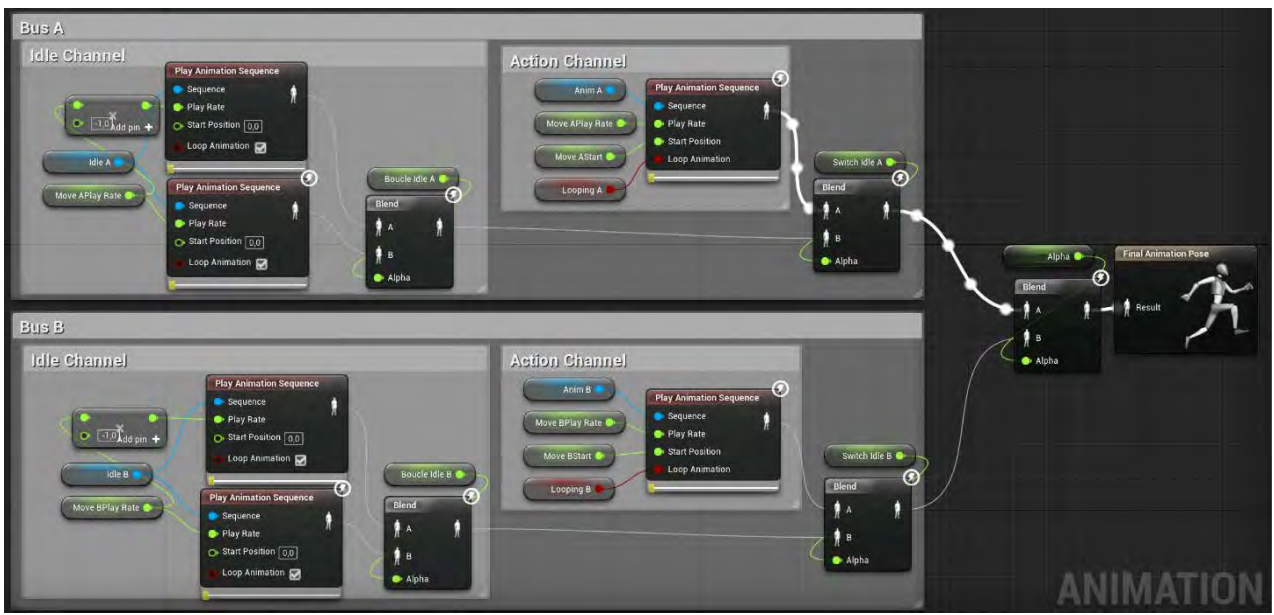

**Figure 3. Unreal Engine 4 animation Finite State Machine of a specific type of OAV**

Firstly, OAV can be controlled in real time by a mocaptor, acting in front of the screen, or aside the stage, hidden or not. The OAV figures out a living shadow that acts on the 3D stage in respect of a scenario in relationships with the actions of its partner on the physical stage. The setup is currently using Noitom's Neuron Motion Capture suit and Axis Neuron software to process the data. The motion retargeting from the Axis Neuron data to the OAV virtual character is done with Autodesk Motion Builder and the Noitom's MotionRobot plugin. The retargeted data are sent to Unreal Engine 4 video game engine (UE4) through the LiveLink plugin, released in 2018 by Epic Game to facilitate on-set previsualization with third-party softwares. Secondly, OAV actions are recorded and combined to block a scenario. However, theater imposes to respect a "real time acting process" condition to keep its essential nature of living performing art. Looking at performers constrained by the imposed rhythm of a time fixed movie is unacceptable for the audience. A solution has been found by programming a specific virtual puppet based on a finite state machine (FSM) mixing two buses, each one combining the same mix of two animation channels (fig. 3). Action channel uses any action starting from one idle pose to another one. Idle channel uses non looping idle animation.

3.3.2. Idle animation built on the shelf
Considering that CAVOAV setup is dedicated to theatrical performances, projects involve often a lot of characters acting during a long time according to animation standards, closer to an hour than few minutes. UE4 offers a tool, called Sequence Recorder, that allows to record animation track of any object during a motion capture session. This tool allowed Epic Game to win the *Real-Time Live!* contest in Siggraph 2016 conference with a demonstration showing a director shooting a short movie featuring two characters, with the same performer, in real time (Antoniades, 2016).
The UE4 Sequence Recorder is used to record "step by step" actions with their starting and final idle poses. The motion capture sessions are done with a mocaptor controlling the OAV living puppet (see 3.1.1.) in the virtual set and following instructions of the stage director, as in traditional theater rehearsals. In a post-production phase, each animation is cut in three parts directly inside UE4: starting idle, action, ending idle (than is also the starting idle of the next action) poses. The starting and ending parts feed the idle channel of the animation FSM.
The idle channel is composed of a sub-system that mixes the animation with itself in a reverse mode in order to avoid the looping jump effect, given that an idle animation never ends with the same pose as it starts. This sub-system allows to quickly blend any animation to an idle one in order to suspend the OAV action, without having to build a proper animation with a third-party software as Motion Builder. It considerably shortens the time to build a complex scenario with numerous OAVs on stage.
Finally, the OAV actions are triggered step by step by an operator following living actions done by the performers on the physical stage. Therefore, the CAVOAV setup offers the minimal conditions to have physical actors and digital characters acting together in a mixed reality environment.

## 4. Two use cases
CAVOAV has been used in two different contexts in 2019. The first one involves pedagogical masterclasses and workshops intended to introduce theater students to act and improvise with avatars on a mixed reality stage. The second one results in the production of a performance, *The Shadow*, that toured in Ukraine and France in fall 2019 (Gagneré, 2019).

### 4.1. Pedagogical use
For example, fig. 4 left image shows a student-mocaptor who controls the OAV in the center of a virtual set representing a room with libraries and fireplace. The shadow of the living OAV is very visible and the student-mocaptor is asked to take it into account in the expressivity of the virtual OAV movements. Another animated OAV is present backstage and will act specific actions with which the living OAV will have to interact.
They are two acting challenges for the mocaptor in this kind of pedagogical situation: inhabiting the virtual body of the OAV and moving properly in the virtual set, making the movement esthetically interesting according to the reinforced quality of presence given by the status of the virtual body shadow. Moreover, students are invited to play in their improvisation with all the 3D visual effects on hand (see 3.2).

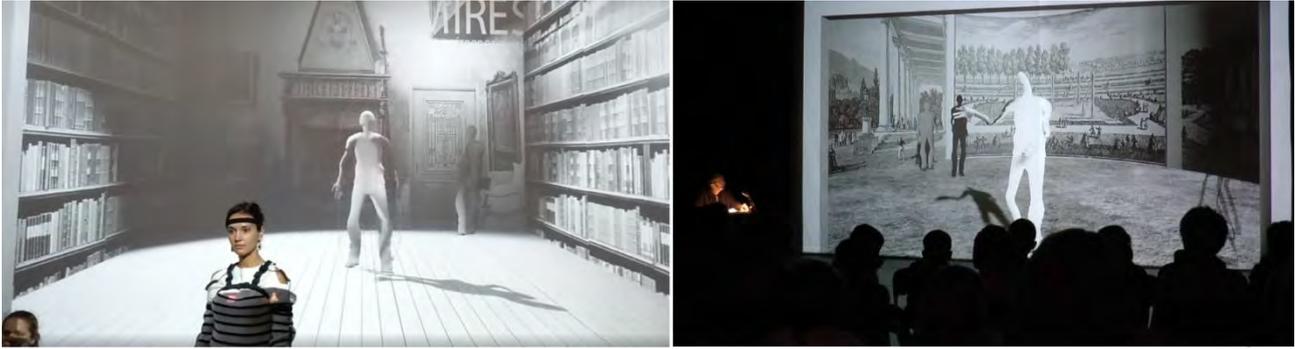

**Figure 4.** Masterclasses (left) and *The Shadow* performance (right) use cases.

Even if an immersive use of CAVOAV has not yet been practiced, the combination of avatar embodiment and play with the avatar own shadow results in an acting constraint that elicits the students to a better understanding of the expressive possibilities of the mixed reality stage virtual part. That deepens the interaction qualities between physical and digital partners. Virtual reality brings a new and unique possibility of making alive 2D shadows and enlarges the range of theatrical situations.

### 4.2. *The Shadow* performance

*The shadow* is a performance after the homonymous Andersen's tale (1847). It tells the story of a Scientist who lost its shadow during a journey abroad and met it again many years later. The Shadow became a sort of human, but unfortunately deprived of its own shadow. He came back to make a deal with the Scientist in order to get married to a Princess.

Fig. 2 and fig. 4 right images show two different moments of a public performance. Fig. 2 is an excerpt of the beginning with the transformation of five "traditional" shadows in five OAV characters. Fig. 4 happens later and shows a theatrical situation involving the Princess, frontstage in white with shadow, and near the cylinder backstage the Shadow in black without shadow and the Scientist in grey with shadow. The performer is sitting left aside at a small table. He tells the story and triggers cues with a midi controller, step by step, according to the text and his way of acting. All the OAV actions have been recorded previously by the director with a mocaptor controlling an OAV living puppet in the different sets. Fig. 4 right image shows a moment of interaction between the performer and the OAV white Princess.

Performances have been given to young children from 8 to 12 years old and some impressions have been informally collected. Globally, audiences were immersed in the story. Some children weren't even conscious of the real time process and perceived the animation as a feature film. Others believed that hidden actors were playing offstage, directly reacting to the actor's words. The majority were very intrigued by the OAV shape and believed it was of paper or aluminum, not aware of motion capture possibilities. CAVOAV gave promising results as an artistic tool to captivate audiences.

### 5. Conclusion and perspectives

In this paper, we introduced a research mixing the theatrical direction of shadow avatars and the building of a virtual theater setup inspired by traditional shadow theater in a mixed reality environment. We described the virtual 3D setup, the nature of the shadow avatars (OAVs) and the issues of directing believable interactions between virtual avatars and physical performers on stage. Two modalities of OAV direction have been exposed. The shadow avatar virtual theater (CAVOAV) has been used in two contexts: the development of theatrical creativity in mixed reality through pedagogical workshops; and an artistic achievement in *The Shadow* performance, after H. C. Andersen.

We intend to develop the research in mixed reality environment by rehearsing more complex interactions between physical performers and OAVs, notably by exploring OAVs virtual golem behavior. We would like also to test CAVOAV setup in an immersive virtual reality context, both for performers and audience.

## 6. Acknowledgments

We would like to specially thank didascalie.net performing arts structure which supports the development of AvatarStaging framework and CAVOAV setup, and produces *The Shadow* (http://didascalie.net).